\documentclass[11pt,preprintnumbers,superscriptaddress,amsmath,amssymb,nofootinbib]{revtex4-1}
\usepackage{graphicx} 
\usepackage{dcolumn}
\usepackage{bm}
\usepackage{amssymb}
\usepackage{amsmath}
\usepackage{epsfig}    
\usepackage{color}
\usepackage{slashed}
\usepackage {threeparttable}
\usepackage[pdftex]{hyperref} 
 
\usepackage{url}
\allowdisplaybreaks[1]

\begin{document} 

\title{H-COUP Version 2:\\ a program for one-loop corrected Higgs boson decays\\
in non-minimal Higgs sectors}


\preprint{OU-HET 1024}
\preprint{KA-TP-17-2019}

\author{Shinya Kanemura}
\email{kanemu@het.phys.sci.osaka-u.ac.jp}
\affiliation{Department of Physics, Osaka University, Toyonaka, Osaka 560-0043, Japan}
\author{Mariko Kikuchi}
\email{kikuchi@kct.ac.jp}
\affiliation{National Institute of Technology, Kitakyushu College, Kitakyushu, Fukuoka 802-0985, Japan}
\author{Kentarou~Mawatari\footnote{Address after October 2019: Faculty of Education, Iwate University, Morioka, Iwate 020-8550, Japan}}
\email{mawatari@iwate-u.ac.jp}
\affiliation{Department of Physics, Osaka University, Toyonaka, Osaka 560-0043, Japan}

\author{Kodai Sakurai}
\email{kodai.sakurai@kit.edu}
\affiliation{Department of Physics, Osaka University, Toyonaka, Osaka 560-0043, Japan}
\affiliation{Institute for Theoretical Physics, Karlsruhe Institute of Technology, 76131 Karlsruhe, Germany}

\author{Kei Yagyu}
\email{yagyu@het.phys.sci.osaka-u.ac.jp}
\affiliation{Department of Physics, Osaka University, Toyonaka, Osaka 560-0043, Japan}

\begin{abstract}

{ We present the concept of {{\tt H-COUP\_ver~2}}, which evaluates {the decay rates} (including higher order corrections) for the Higgs boson with a mass of 125{\rm GeV} in various extended Higgs models.}
In the previous version ({\tt H-COUP\_1.0}), only a full set of the Higgs boson vertices are evaluated at one-loop level 
in a gauge invariant manner in these models. 
{{\tt H-COUP\_ver~2}} contains all the functions of {\tt H-COUP\_1.0}. 
After shortly introducing these extended Higgs models and discussing their theoretical and experimental constraints, 
we summarize formulae for the renormalized vertices and the decay rates. 
We then explain how to install and run {{\tt H-COUP\_ver~2}} with some numerical examples.  

\end{abstract}
\maketitle

\section*{New version  Program summary}
\hspace{-14.4pt}{\bf Program title}: H-COUP Version 2
\vspace{9.6pt}

\hspace{-14.4pt}{\bf Licensing provisions}: GPLv3
\vspace{9.6pt}

\hspace{-14.4pt}{\bf Programming language}: Fortran90
\vspace{9.6pt}

\hspace{-14.4pt}{\bf Program obtained from}: \url{http://www-het.phys.sci.osaka-u.ac.jp/~hcoup/}
\vspace{9.6pt}

\hspace{-14.4pt}{\bf Journal Reference of previous version}: \url{https://doi.org/10.1016/j.cpc.2018.06.012}
\vspace{9.6pt}

\hspace{-14.4pt}{\bf Does the new version supersede the previous version?}: Yes
\vspace{9.6pt}

\hspace{-14.4pt}{\bf Reasons for the new version}: The previous version ({\tt H-COUP\_1.0}), which numerically evaluates the full set of one-loop corrected vertex functions for the Higgs boson with a mass of 125 GeV, does not automatically provide the decay rates, the total decay width and branching ratios. 
In addition, QCD corrections have not been included in the previous version. 

\vspace{9.6pt}

\hspace{-14.4pt}{\bf Summary of revisions}: Decay branching ratios and the total decay width for the Higgs boson with NLO electroweak and NNLO QCD corrections are added as outputs. 

\vspace{9.6pt}

\hspace{-14.4pt}{\bf Nature of problem}: Decay rates for the Higgs boson and the total decay width are numerically evaluated at NLO for electroweak and NNLO for QCD in the Higgs singlet model, four types (Type-I, Type-II,  Type-X, Type-Y) of two Higgs doublet models with a softly-broken $Z_2$ symmetry and the inert doublet model. 
\vspace{9.6pt}\vspace{9.6pt}

\hspace{-14.4pt}{\bf Solution of method}: 
Electroweak and QCD corrections to the decay rates are computed by the improved on-shell scheme and the $\overline{\rm MS}$ scheme, respectively. 

\vspace{9.6pt}

\hspace{-14.4pt}{\bf Additional comments including restrictions and unusual features}: All functions of the previous version are included in   {{\tt H-COUP\_ver~2}}. 
\vspace{9.6pt}

\clearpage
\hspace{-14.4pt}{\bf References}:
          
     \begin{enumerate}
     \renewcommand{\labelenumi}{[\arabic{enumi}]}
\item S. Kanemura, M. Kikuchi, K. Sakurai, K. Yagyu, Comput. Phys. Commun. {\bf 233} (2018) 134-144 [arXiv:1710.04603 [hep-ph]].
\item S. Kanemura, M. Kikuchi, K. Mawatari, K. Sakurai, K. Yagyu, Nucl. Phys. B {\bf 949} (2019) 114791 [arXiv:1906.10070 [hep-ph]].
\item S. Kanemura, M. Kikuchi, K. Mawatari, K. Sakurai, K. Yagyu, Phys. Lett. B {\bf 783} (2018) 140-149 [arXiv:1803.01456 [hep-ph]].
\item S. Kanemura, M. Kikuchi, K. Sakurai, K. Yagyu, Phys. Rev. D {\bf 96} (2017) 035014 [arXiv:1705.05399[hep-ph]].
\item S. Kanemura, M. Kikuchi, K. Yagyu, Nucl. Phys. B {\bf 907} (2016) 286-322 [arXiv:1511.06211 [hep-ph]].
\item S. Kanemura, M. Kikuchi, K. Yagyu, Nucl. Phys. B {\bf 917} (2017) 154-177 [arXiv:1608.01582 [hep-ph]].
\item S. Kanemura, Y. Okada, E. Senaha, C. Yuan, Phys. Rev. D {\bf 70} (2004) 115002 [hep-ph/0408364]. 
\item S. Kanemura, M. Kikuchi, K. Yagyu, Phys. Lett. B {\bf 731} (2014) 27-35 [arXiv:1401.0515 [hep-ph]].
\item S. Kanemura, M. Kikuchi, K. Yagyu, Nucl. Phys. B {\bf 896} (2015) 80-137 [arXiv:1502.07716 [hep-ph]]. 
\item S. Kanemura, M. Kikuchi, K. Sakurai, Phys. Rev. D {\bf 94} (2016) 115011 [arXiv:1605.08520 [hep-ph]]. 
   \end{enumerate}

\newpage

\section{Introduction}

By the discovery of the Higgs boson at the LHC, exploring details of the scalar sector, which is responsible for the electroweak (EW) symmetry breaking, 
has become one of the most important subjects of high energy particle physics. 
The current situation clarified by collider experiments can be summarized by two important things, 
(i) properties of the discovered Higgs boson are consistent with those of the Standard Model (SM) Higgs boson within theoretical and experimental uncertainties, and (ii)
other new particles have not yet been observed. 
These current experimental results can be explained within the minimal Higgs sector assumed in the SM, which is composed of one isospin scalar doublet. 

On the other hand, the Higgs sector of the SM  does not have any principle to determine its structure, differently from the gauge sector. 
Thus, non-minimal forms of the Higgs sector should also be considered as well, unless they are excluded by the current data.  
In addition, non-minimal Higgs sectors are often predicted in various new physics models which have been proposed to solve problems of the SM; 
i.e., the hierarchy problem as well as the existence of phenomena which cannot be explained in the SM such as 
neutrino oscillations, dark matter and baryon asymmetry of the Universe. 
Because the structure of the Higgs sector can strongly depend on each new physics scenario, exploring the shape of the Higgs sector is a key 
to determine the direction of new physics beyond the SM. 

When additional Higgs bosons are discovered at future collider experiments, it provides direct evidence for a non-minimal Higgs sector. 
The structure of the non-minimal Higgs sector is then narrowed down by properties of additional Higgs bosons; e.g., electric charges, masses, couplings and so on. 
However, even if the second Higgs boson is not directly discovered, we can indirectly determine the structure of the Higgs sector by measuring deviations in observables for the discovered Higgs boson from the SM predictions such as 
couplings, the width, branching ratios and production cross sections.   
Currently, these observables are not sufficiently measured with enough accuracy for indirect searches for additional Higgs bosons~{\cite{Sirunyan:2018koj,Aad:2019mbh}}. However,  
they are expected to be precisely measured at future experiments, such as the high-luminosity LHC (HL-LHC)~{\cite{CMS:2013xfa,ATLAS:2013hta,Cepeda:2019klc}}, the International Linear Collider (ILC)~\cite{Baer:2013cma,Fujii:2017vwa,Asai:2017pwp,Fujii:2019zll}, 
the Future Circular Collider (FCC)~\cite{Gomez-Ceballos:2013zzn}, 
the Circular Electron Positron Collider (CEPC)\cite{CEPC-SPPCStudyGroup:2015csa} and 
the Compact LInear Collider (CLIC)~\cite{CLIC:2016zwp}.
{Measurements of the Higgs boson properties at these future colliders have been summarized in Ref.~\cite{Strategy:2019vxc}. }
For example, couplings of the discovered Higgs boson are expected to be measured with one percent level or better at the ILC with the center of mass energy of 250 GeV. 
Therefore, accurate calculations of the Higgs boson observables with radiative corrections are necessary to compare with their precisely measured values. 

\begin{table}[t]\begin{center}
\renewcommand{\thempfootnote}{\arabic{mpfootnote}}
\begin{tabular}{c|cccccccc|ccc}\hline\hline
               & {\footnotesize $hf\bar{f}$}  &{\footnotesize $hVV$} & {\footnotesize $hhh$} &{\footnotesize $\Gamma(h\to f\bar{f})$\ [QCD]} &{\footnotesize $\Gamma(h\to f\bar{f})$\ [EW]}&{\footnotesize $\Gamma(h\to VV)$}   \\\hline
SM             &\cite{Fleischer:1980ub,Kniehl:1991ze,Dabelstein:1991ky,Kniehl:1993ay}&\cite{Fleischer:1980ub,Kniehl:1990mq,Kniehl:1991xe,Kniehl:1993ay}& \cite{Kanemura:2002vm,Kanemura:2004mg} &\cite{Drees:1989du,Braaten:1980yq,Sakai:1980fa,Inami:1980qp,Kniehl:1993ay} & \cite{Fleischer:1980ub,Kniehl:1991ze,Dabelstein:1991ky,Kniehl:1993ay}& {\cite{Fleischer:1980ub,Kniehl:1990mq,Kniehl:1991xe,Kniehl:1993ay,Bredenstein:2006rh,Bredenstein:2006nk,Bredenstein:2006ha}}\\\hline
MSSM          &\cite{Dabelstein:1995js,Heinemeyer:2001qd,Frank:2006yh} & {\cite{Chankowski:1992er,Hahn:2002gm,Heinemeyer:2001qd,Frank:2006yh,Heinemeyer:2015qbu,Williams:2011bu}} & \cite{Hollik:2001px,Dobado:2002jz,Williams:2007dc,Dolgopolov:2003kv} &{\cite{Drees:1989du,Chankowski:1992er,Haber:2000kq,Heinemeyer:2001qd,Frank:2006yh,Heinemeyer:2000fa,Williams:2011bu} }& {\cite{Dabelstein:1995js,Heinemeyer:2001qd,Frank:2006yh,Heinemeyer:2000fa,Williams:2011bu} }& {\cite{Hollik:2010ji,Hollik:2011xd,Gonzalez:2012mq,Heinemeyer:2001qd,Frank:2006yh} }  \\\hline
{NMSSM} & & & {\cite{Nhung:2013lpa}} &{\cite{Baglio:2013vya,Domingo:2018uim,Baglio:2019nlc}}&{\cite{Domingo:2018uim,Baglio:2019nlc}}&{\cite{Domingo:2018uim,Baglio:2019nlc}} \\\hline
THDMs              & {\cite{Arhrib:2003ph,Arhrib:2016snv, Kanemura:2014dja, Kanemura:2015mxa,Kanemura:2017wtm,Chen:2018shg,Gu:2017ckc} }     &  {\cite{Kanemura:2004mg, LopezVal:2010vk,Castilla-Valdez:2015sng,Kanemura:2015mxa, Altenkamp:2017ldc, Kanemura:2017wtm,Chen:2018shg,Gu:2017ckc,Xie:2018yiv} } &\cite{Kanemura:2004mg, Kanemura:2015mxa, Kanemura:2017wtm} & &{\cite{Arhrib:2003ph,Arhrib:2016snv,Kanemura:2018yai,Kanemura:2019kjg,Xie:2018yiv}} & \cite{Castilla-Valdez:2015sng, Altenkamp:2017ldc,Kanemura:2018yai,Kanemura:2019kjg}   \\\hline
HSM               & \cite{Kanemura:2015fra,Kanemura:2017wtm}  & \cite{Kanemura:2015fra,Kanemura:2017wtm} & \cite{Kanemura:2016lkz, He:2016sqr,Kanemura:2017wtm} &&\cite{Kanemura:2018yai,Kanemura:2019kjg}&\cite{Kanemura:2018yai,Kanemura:2019kjg}\\\hline 
IDM               & \cite{Kanemura:2016sos,Kanemura:2017wtm}          & \cite{Arhrib:2015hoa,Kanemura:2016sos,Kanemura:2017wtm} & \cite{Arhrib:2015hoa,Kanemura:2016sos,Kanemura:2017wtm} &&\cite{Kanemura:2019kjg}& \cite{Kanemura:2019kjg}\\\hline\hline
\end{tabular}
\caption{Summary for studies on radiative corrections to the Higgs boson couplings at one-loop level as well as Higgs boson decay rate including at next-to-leading order (NLO). 
For the $h\to f\bar{f}$ , we separately show the works for the NLO QCD corrections 
and EW corrections. 
}
\label{tab:previous}
\end{center}
\end{table}

There have been many studies on radiative corrections to the vertex functions and decay rates of the Higgs boson $h$(125) in various non-minimal Higgs sectors and new physics models in addition to the SM, 
where $h$(125) represents the discovered Higgs boson with the mass of 125 GeV. 
In Table \ref{tab:previous}, we summarize previous studies on one-loop corrections to the $hf\bar{f}$, $hVV$ ($V = W,Z$) and $hhh$ vertices as well as the decay rates for $h\to f\bar{f}$, $h\to VV$ in the SM, MSSM, {NMSSM, }two Higgs doublet models (THDMs) with a softly broken $Z_2$ symmetry, the Higgs singlet model (HSM) and the inert doublet model (IDM). 
One can numerically evaluate these vertex functions and decay rates with higher order corrections by using several public tools. 
For the SM and MSSM (next to MSSM), {\tt HDECAY}~\cite{Djouadi:1997yw,Djouadi:2018xqq}, {\tt FeynHiggs }~{\cite{Heinemeyer:1998yj,Hahn:2009zz,Hahn:2010te,Bahl:2018qog}} and {\tt HFOLD}\cite{Frisch:2010gw} ({\tt NMHDECAY}~\cite{Ellwanger:2004xm,Ellwanger:2005dv}, {\tt NMSSMCALC}~\cite{Baglio:2013iia} {and {\tt NMSSMCALCEW}~\cite{Baglio:2019nlc}}) can compute decay width and branching ratios of Higgs bosons with EW corrections and QCD corrections.
Regarding the extended Higgs models,  {\tt 2HDMC}~\cite{Eriksson:2009ws} and {\tt sHDECAY}~\cite{Costa:2015llh} can give decay rates and total width, and branching ratios of Higgs bosons with QCD corrections in THDMs and the HSM. Also, {\tt 2HDECAY}~\cite{Krause:2018wmo} can provide the decay rates and branching ratios with both EW corrections and QCD corrections in THDMs. 
Apart from these public tools, as a first tool to observables for $h$(125) with one-loop EW corrections in various non-SUSY models with extended Higgs sectors, {\tt H-COUP\_1.0}~\cite{Kanemura:2017gbi} had been published. 

In this article for the manual,
we present {{\tt H-COUP\_ver~2}}, a fortran program for numerical evaluation of decay rates of the discovered Higgs boson $h(125)$ 
with next-to-leading order (NLO) EW and scalar loop corrections, and next-to-next-to leading order (NNLO) QCD corrections in a variety of extended Higgs models
such as the HSM, four types (Type-I, Type-II, Type-X, Type-Y) of THDMs with a softly-broken $Z_2$ symmetry, and the IDM. 
{ {{\tt H-COUP\_ver~2}} {can also} evaluate decay rates of $h(125)$ in the SM with the same accuracy. 
We have confirmed  that numerical values for the decay rates of $h \to f\bar{f}$ in the THDMs with the EW and scalar loop corrections at NLO  
are in good agreement with those computed by {\tt 2HDECAY}. } 
In the previous version ({\tt H-COUP\_1.0}), a full set of vertices for $h$(125) can be evaluated at one-loop level in 
the improved on-shell scheme for NLO EW in these models. 
By extending the {\tt H-COUP\_1.0} functionalities, we completed the calculations of all the decay rates of $h$(125) as {{\tt H-COUP\_ver~2}}.
Therefore, {{\tt H-COUP\_ver~2}} contains all the functions of {\tt H-COUP\_1.0}.

Physics results obtained by preliminary version of {{\tt H-COUP\_ver~2}} have been presented in Refs.~\cite{Kanemura:2018yai,Kanemura:2019kjg} where NLO EW and NLO QCD corrections were implemented. 
We note that, with a process to make a public version of the {{\tt H-COUP\_ver~2}} program, 
we added NNLO-QCD corrections to the $h\to q\bar q$, $gg$, $\gamma\gamma$ modes. 
We also added $h\to\mu\mu$ for the completeness of the list of the decay modes. 


This article is organized as follows. 
In Sec.~\ref{sec:model}, we briefly review the extended Higgs models, and define input parameters for each model. 
In Sec.~\ref{sec:decay}, we discuss renormalized vertices and decay rates of $h$(125) based on Refs.\cite{Kanemura:2017wtm,Kanemura:2015fra,Kanemura:2016lkz,Kanemura:2004mg,Kanemura:2014dja,Kanemura:2015mxa,Kanemura:2016sos,Kanemura:2018yai,Kanemura:2019kjg} which are implemented in {{\tt H-COUP\_ver~2}}. 
In Sec.~\ref{sec:structure}, the structure of {{\tt H-COUP\_ver~2}} is explained. 
In Sec.~\ref{sec:how}, the installation and how to run {{\tt H-COUP\_ver~2}} are described with some numerical examples. 
Summary of this manual is given in Sec.~\ref{sec:summary}. 

\section{Models and constraints}\label{sec:model}

In this section, we define the Higgs sectors of the HSM, the THDMs { with the CP-conservation} and the IDM.  
In particular, we uniformly and compactly introduce mass eigenstates of the scalar fields and  
free input parameters in each model. 
Since all the models are exactly the same as those in {\tt H-COUP\_1.0}, see the manual of Ver.1~\cite{Kanemura:2017gbi} for details of definitions and descriptions about the models such as Lagrangian and some formulae.

In the all models covered in this manual, 
mass eigenstates of scalar fields are commonly represented as follows; 
\begin{align}
 h \space{}&:\space{}\textrm{the discovered CP-even Higgs boson with the mass 125 GeV,} \notag\\
 H \space{}&:\space{}\textrm{another CP-even Higgs boson, } \notag\\
 A \space{}&:\space{}\textrm{a CP-odd Higgs boson, }\\
 H^\pm \space{}&:\space{}\textrm{a pair of singly charged Higgs bosons}. \notag
\end{align} 
{In the following descriptions for each extended Higgs model, a mass of $h$ is always set {to} $m_h=125\ {\rm GeV}$, so that the parameter $m_h$ does not appear in inputs parameters for each model which is stored as {the} SM input parameters in {the} {\tt H-COUP}  program.} 
{{\tt H-COUP\_ver~2}} incorporates some theoretical constraints,  
i.e., the tree-level unitarity bound, the triviality bound, the vacuum stability bound (tree level and improved by renormalization group equations (RGEs) ) and the true vacuum condition, as well as an experimental constraint by the EW
S and T parameters, which are exactly the same as those in {\tt H-COUP\_1.0}. 
Detailed descriptions for the constraints are given in the manual of {\tt H-COUP\_1.0}~\cite{Kanemura:2017gbi}. 
{We note that {{\tt H-COUP\_ver~2}} gives values of the branching ratios and the width of h(125), even if one of constraints are not fulfilled, while a warning appears to indicate the constraints which 
are not satisfied.}

\subsection{HSM}
The Higgs sector of the HSM is composed of the SM Higgs field $\Phi$, i.e., the isospin doublet Higgs field with hypercharge $Y=1/2$, and an isospin singlet scalar field $S$ with $Y=0$. 
Detailed definitions of descriptions about the HSM are given 
in Refs.~\cite{Kanemura:2015fra,Kanemura:2016lkz},  
whose notation is the same as the notation in this article.  
After the EW symmetry breaking, there appear two physical scalar states $h$ and $H$ by the mixing of neutral components of $\Phi$ and $S$.  
The Higgs potential has 8 free parameters. 
Two of them, the mass of $h$(125) $m_h^{}$, and the vacuum expectation value (VEV) $v$ of the doublet field $\Phi$  
are fixed, i.e., $m_h^{} = 125$ GeV and $v\simeq 246$ GeV. 
Moreover, the VEV of the singlet field, can be absorbed by the field redefinition~\cite{Chen:2014ask}.  
Here, the following 5 parameters are chosen as input free parameters; 
\begin{align}
 m_H,\; \alpha,\; \lambda_S^{},\; \lambda_{\Phi S},\; \mu_S^{},  
\end{align} 
where $m_H^{}$ and $\alpha$ are the mass of $H$ and the mixing angle between $h$ and $H$, respectively. 
We define the range of $\alpha$ as $-\pi/2 \leq \alpha \leq \pi/2$.   
The remaining three parameters are the original parameters given in the potential.

\subsection{THDMs}
THDMs contain two isospin doublet Higgs fields $\Phi_1^{}$ and $\Phi_2^{}$ with $Y=1/2$. 
In these models with a softly broken $Z_2^{}$ symmetry, 
the two scalar fields are assigned different $Z_2^{}$ charges with each other. 
{{\tt H-COUP\_ver~2}} covers four types of {CP-conserving} THDMs with different Yukawa interactions~\cite{Barger:1989fj, Grossman:1994jb,Aoki:2009ha}, 
which are called Type-I, Type-II, Type-X and Type-Y.  
Please see Refs.~\cite{Kanemura:2004mg, Kanemura:2014dja, Kanemura:2015mxa} for details of the models.  
In the THDMs, three neutral scalar fields ($h, H$ and $A$) and a pair of singly charged scalar fields ($H^\pm$) appear as mass eigenstates.  
We choose the following 6 parameters as input free parameters,  
\begin{align}
 m_H^{},\; m_A^{},\; m_{H^\pm}^{},\; \sin(\beta-\alpha),\; \tan\beta,\; M^2, 
\end{align} 
where $\sin(\beta-\alpha) \geq 0$ and $\tan\beta > 0$ are taken, 
$ m_H^{}, m_A^{}, m_{H^\pm}^{}$ represent masses of the additional Higgs bosons, and 
$\alpha$ ($\beta$) is a mixing angle of CP-even (CP-odd) scalar components, and
$M^2$ is a parameters describing the soft breaking scale of the $Z_2^{}$ symmetry. 
When we take $\sin(\beta-\alpha)$ and $\tan\beta$ as input parameters, 
we also have to specify the sign of $\cos(\beta-\alpha)$.

\subsection{IDM}
The Higgs sector of the IDM consists of two isospin doublet Higgs fields $\Phi$ and $\eta$ with $Y=1/2$. 
This model has unbroken $Z_2^{}$ symmetry, so that 
$\Phi$ and $\eta$ with different $Z_2^{}$ charges do not mix their components. 
As a result, there are five types of scalar particles, i.e., $h$, $H$, $A$ and $H^\pm$,  
where $h$ ($H$, $A$ and $H^\pm$) is the original component of $\Phi$ ($\eta$). 
In {\tt H-COUP}, the following five parameters, 
\begin{align}
 m_H^{},\; m_A^{},\; m_{H^\pm}^{},\; \mu_2^{},\; \lambda_2^{}, 
\end{align}
are taken as for input parameters, where  
$\mu_2^{}$ and $\lambda_2^{}$ are coefficient parameters of the quadratic and quartic terms of $\eta$ in the potential, respectively.  
Details of definitions, formulae and descriptions for the IDM are given in Refs.~\cite{Kanemura:2016sos}.  

\section{Renormalized vertices and decay rates\label{sec:decay}}

In this section, renormalized vertex functions for the discovered Higgs boson $hf\bar{f}$, $hVV$ ($V=W\, {\rm or}\, Z$) and $hhh$ at one-loop are defined. 
Subsequently, analytical expressions of the decay rates with higher order corrections are described, i.e., $h\to f\bar{f},{\ h\to VV^\ast \to Vf\bar{f}},\ h\to gg$ and  $h\to \mathcal{V} \gamma$ $(\mathcal{V}= \gamma\ {\rm or}\ Z)$. 
These quantities are output parameters of the {{\tt H-COUP\_ver~2}}. 
{In {{\tt H-COUP\_ver~2}}, the decays into extra Higgs bosons, e.g., $h\to HH $, $h\to AA $, are not contained.}  

Here, we outline the renormalization scheme for calculations of radiative corrections in {\tt H-COUP}. 
All Feynman diagrams are computed in the 't Hooft-Feynman gauge, and the UV divergences are renormalized by applying the improved on-shell scheme~\cite{Kanemura:2017wtm} for EW corrections. {``Improved" means} the gauge dependence arising from a mixing of scalar fields, {e.g., $H$-$h$ and $G^0$-$A$ for THDMs,} is got rid of by utilizing the pinch technique~\cite{Bojarski:2015kra,Krause:2016oke,Kanemura:2017wtm}\footnote{
{It was found that the implemented renormalization scheme without the gauge dependence given in Ref.~\cite{Kanemura:2017wtm} is necessary to be modified for the treatment of the counterterms for the mixing parameters. 
In Ref.~\cite{Kanemura:2017wtm}, the pinch-terms, which are needed to realize the gauge independent counterterms, are introduced to not only these counterterms originated from the shift of the couplings but also those from the shift of the scalar fields. 
However, the latter should be defined without the pinch-terms. 
In {\tt H-COUP\_ver~2}, this problem was corrected. }
}. 
{In this scheme, all the physical Higgs boson masses and mixing angles are determined by on-shell conditions.} 
On the other hand, for the NLO and NNLO QCD corrections to the Higgs decay processes, the ${\rm \overline{MS}}$ scheme is applied. 

Apart from the UV divergences, for the decay of $h\to f\bar{f}$ and $\ h\to VV^\ast$ and also $hf\bar{f}$ and $hVV$ vertex functions, IR divergences appear in Feynman diagrams with a virtual photon, which are cancelled with contributions from real photon emission. 
For the decay of $h\to f\bar{f}$ and $h\to ZZ^\ast$, virtual photon loop corrections can be separated  from weak corrections and analytical formulae for the total corrections (virtual photon loop corrections plus real photon emissions) have already known in the SM. 
Since these QED corrections at the NLO for extended Higgs models are common with those of the SM, the analytical formulae of the total QED corrections are  simply implemented in {{\tt H-COUP\_ver~2}}. 
Related to these treatment of the QED corrections to $h\to f\bar{f}$ and $h\to VV^\ast$, virtual photon loop corrections are switched off  in evaluations of the $hf\bar{f}$ and $hZZ$ vertex functions. 
On the other hand, photon loop corrections and weak corrections to the ${h\to WW^\ast \to Wf\bar{f^{\prime}}}$ are not separable. 
Therefore, virtual photon corrections  and contributions of real photon emissions are individually evaluated. 
The latter is evaluated by using the phase space slicing method~\cite{Harris:2001sx}, thus photon phase space is divided into the soft region and the hard region. 
While the analytical expressions are implemented in {\tt H-COUP} for contributions with soft photon, the numerical values for contributions with hard photon are evaluated by {\tt Madgraph5\_aMC@NLO}~\cite{Alwall:2014hca} with default values for SM parameters in {{\tt H-COUP\_ver~2}}. 

\subsection{Renormalized vertex functions}
The renormalized $hf\bar{f}$ and  $hVV$ vertices are expressed  in terms of form factors as
\begin{align}
\hat{\Gamma}_{h ff}(p_1^2,p_2^2,q^2)&=
\hat{\Gamma}_{h ff}^S+\gamma_5 \hat{\Gamma}_{h ff}^P+p_1\hspace{-3.5mm}/\hspace{2mm}\hat{\Gamma}_{h ff}^{V_1}
+p_2\hspace{-3.5mm}/\hspace{2mm}\hat{\Gamma}_{h ff}^{V_2}\notag\\
&\quad +p_1\hspace{-3.5mm}/\hspace{2mm}\gamma_5 \hat{\Gamma}_{h ff}^{A_1}
+p_2\hspace{-3.5mm}/\hspace{2mm}\gamma_5\hat{\Gamma}_{h ff}^{A_2}
+p_1\hspace{-3.5mm}/\hspace{2mm}p_2\hspace{-3.5mm}/\hspace{2mm}\hat{\Gamma}_{h ff}^{T}
+p_1\hspace{-3.5mm}/\hspace{2mm}p_2\hspace{-3.5mm}/\hspace{2mm}\gamma_5\hat{\Gamma}_{h ff}^{PT}, \label{eq:hff-form} \\
\hat{\Gamma}_{hVV}^{\mu\nu}(p_1^2,p_2^2,q^2)&=g^{\mu\nu}\hat{\Gamma}_{h VV}^1 + \frac{p_1^\nu p_2^\mu}{m_V^2}\hat{\Gamma}_{h VV}^2 + i\epsilon^{\mu\nu\rho\sigma}\frac{p_{1\rho} p_{2\sigma}}{m_V^2}\hat{\Gamma}_{h VV}^3,  
\end{align}
where $p_1^\mu$ and $p_2^\nu$ for the $hff\ (hVV)$ vertex are defined as incoming momenta of fermion and anti-fermion (two weak gauge bosons), and $q^\mu$ denotes the outgoing momentum of the Higgs boson.
In contrast to these vertices, the $hhh$ vertex $\hat{\Gamma}_{h hh}(p_1^2,p_2^2,q^2)$ is a scalar function. The renormalized scalar functions $\hat{\Gamma}_{hXX}$ are commonly  
divided into  two parts
\begin{align}
\hat{\Gamma}^i_{hXX}(p_1^2,p_2^2,q^2)&=\Gamma^{i,{\rm tree}}_{hXX} + \Gamma^{i,{\rm loop}}_{hXX}(p_1^2,p_2^2,q^2),  \label{eq:form-loop}
\end{align}
where the loop part $\Gamma^{i,{\rm loop}}_{hXX}$ is further decomposed into 1PI diagram contributions and counterterm contributions, i.e. $\Gamma^{i,{\rm loop}}_{hXX}=\Gamma^{i,{\rm 1PI}}_{hXX}+\delta\Gamma^{i}_{hXX}$. 
The tree-level contributions for each vertex function are written by 
\begin{align}
\Gamma_{hf\bar{f}}^{S,{\rm tree}}=-\frac{m_f}{v}\kappa_f,\ \ \ \ \ \ \ \ \Gamma_{hVV}^{1,{\rm tree}}=\frac{2m^2_V}{v}\kappa_V,\ \ \ \ \ \ \ \ \Gamma_{hhh}^{{\rm tree}}=-\frac{3m^2_h}{v}\kappa_h,\ \ \ 
\end{align}
where scaling factors $\kappa_X$ for each extended Higgs model are summarized in Table~\ref{tab:kappa}, and other form factors become zero at tree level, namely $\Gamma_{hVV}^{2,{\rm tree}}=\Gamma_{hVV}^{3,{\rm tree}}=\Gamma_{hf\bar{f}}^{a,{\rm tree}}=0\ (a\neq S)$. Explicit formula for the loop contributions of each vertex are give in Refs.~\cite{Kanemura:2015fra,Kanemura:2017wtm}, Refs.~\cite{Kanemura:2015mxa,Kanemura:2016lkz,Kanemura:2017wtm}, and Ref.~\cite{Kanemura:2016sos} for the HSM, THDMs and the IDM, respectively.

\begin{table}[t]
\begin{center}
\begin{tabular}{c|ccccccc|}\hline\hline
 & $\kappa_f$ & $\kappa_V$ & $\kappa_h$  \\\hline
HSM        & $c_\alpha$ &$c_\alpha$ & $c_\alpha^3+2s_\alpha^2\frac{v^2}{m_h^2}(c_\alpha\lambda_{\Phi S}-s_\alpha\frac{\mu_S}{v})$\\\hline
THDMs    & $s_{\beta-\alpha} + \zeta_f c_{\beta-\alpha}$  &\ \ \ \ \ \ $s_{\beta-\alpha}$\ \ \ \ \ \  & $s_{\beta-\alpha}+\left(1-\frac{M^2}{m_h^2}\right)c_{\beta-\alpha}^2\left\{2s_{\beta-\alpha}+c_{\beta-\alpha}(\frac{1}{t_\beta}+t_\beta)\right\}$ \\\hline
IDM         & 1 &1 & 1 \\\hline\hline
\end{tabular}
\caption{Scaling factors for Higgs couplings  in the extended Higgs models at tree level. 
The factor $\zeta_f$ in the THDMs varies in accordance with structure of Yukawa interactions , which  is given in Table~\ref{tab:zeta}. }
\label{tab:kappa}
\end{center}
\end{table}
\begin{table}[t]
\begin{center}
{\renewcommand\arraystretch{1.2}
\begin{tabular}{l|ccc}\hline\hline
THDMs&$\zeta_u$ &$\zeta_d$&$\zeta_e$ \\\hline
Type-I&$\cot\beta$&$\cot\beta$&$\cot\beta$ \\\hline
Type-II&$\cot\beta$&$-\tan\beta$&$-\tan\beta$ \\\hline
Type-X (lepton specific)&$\cot\beta$&$\cot\beta$&$-\tan\beta$ \\\hline
Type-Y (flipped) &$\cot\beta$&$-\tan\beta$&$\cot\beta$ \\\hline\hline
\end{tabular}}
\caption{ 
The $\zeta_f$ ($f=u,d,e$) factors appearing in Table~\ref{tab:kappa}. 
}
\label{tab:zeta}
\end{center}
\end{table}

\subsection{Higgs decay rates\label{sec:decay-rate}}

The decay rates for $h\to f\bar{f}$ with higher order corrections can be schematically described as
\begin{align}
\Gamma(h\to f\bar{f})=\Gamma_0(h\to f\bar{f})\left[1+\Delta_{\rm EW}^f+\Delta_{\rm QCD}^f\right],
\end{align}
where $\Gamma_0$ denotes the formula at the LO, i.e.,
\begin{align}
\Gamma_0(h\to f\bar{f}) = \frac{N_c^f}{8\pi}m_h(\Gamma_{hff}^{S,\text{tree}})^2\left(1-\frac{4m_f^2}{m_h^2}\right)^{3/2}, \label{eq:hff-lo}
\end{align}
with $N_c^f=3(1)$ for quark (lepton), and $\Delta_{\rm EW}^f$ and $\Delta_{\rm QCD}^f$ denote the EW corrections and the QCD corrections to $h\to ff$, respectively. 
Hereafter, for all decay modes of the Higgs bosons, we commonly denote the contributions of EW (QCD) corrections as $\Delta^X_{\rm EW (QCD)}$. 
The EW corrections $\Delta_{\rm EW}^f$ at the NLO  can be further divided into the QED corrections (radiative corrections of a photon)  and weak corrections (all the other EW loop corrections) as $\Delta_{\rm EW}^f=\Delta_{\rm QED}^f+\Delta_{\rm Weak}^f$. For the decay into leptons $f=\ell$, the NLO QED correction $\Delta_{\rm QED}^\ell$ is given in the on-shell scheme by~\cite{Kniehl:1991ze,Dabelstein:1991ky,Bardin:1990zj}
\begin{align}
\Delta_{\text{QED}}^\ell =  \frac{\alpha_{\text{em}}}{\pi}Q_\ell^2\left(\frac{9}{4} + \frac{3}{2}\log\frac{m_\ell^2}{m_h^2} \right), \label{eq:del_qed_f}
\end{align} 
and for the decay into quarks, the NLO QED corrections is given in ${\rm \overline{MS}}$ scheme by~\cite{Mihaila:2015lwa} 
\begin{align}
\Delta_{\text{QED}}^q =  \frac{\alpha_{\text{em}}}{\pi}Q_q^2\left(\frac{17}{4}  +\frac{3}{2}\log\frac{\mu^2}{m_h^2} \right),  
\end{align}
where $\mu$ is taken to be $m_h^{}$. 
Whereas, the weak corrections can be commonly expressed in terms of the renormalized Higgs vertex functions as
\begin{align}
\Delta_{\text{weak}}^f =  \frac{2}{\Gamma_{hff}^{S,{\rm tree}}}{\rm Re}\left\{\left[\Gamma_{hff}^{S,{\rm loop}}
+2m_f\Gamma^{V_1,{\rm loop}}_{hff} 
+m_h^2\left(1-\frac{m_f^2}{m_h^2}\right)\Gamma_{hff}^{T,{\rm loop}}\right](m_f^2,m_f^2,m_h^2)\right\} - \Delta r , \label{eq:delew}
\end{align} 
where $\Delta r$ denotes the radiative correction to  muon decay~\cite{Sirlin:1980nh}. 
{It is given by
\begin{align}
\label{eq:defdr}
\Delta r =\frac{{\rm Re}\hat{\Pi}_{WW}(0)}{m_W^2}
+\frac{\alpha_{\rm em}}{4\pi s_W^2} \left(6+\frac{7-4s_W^2}{2s_W^2}\log c^2_W \right),
\end{align}
where $\hat{\Pi}_{WW}$ is the renormalized two-point function for a $W$ boson and the  second term comes from the vertex corrections and box diagram corrections to  muon decay. 
\footnote{{{ 
{By substituting} concrete expressions of $\hat{\Pi}_{WW}$ in Eq.~\eqref{eq:defdr}, 
$\Delta r$ can be written as~{\cite{Kanemura:2015mxa}}
\begin{align}
\notag
\label{eq:dr}
\Delta r &=\Pi^{\rm 1PI}_{\gamma\gamma}(0)^{\prime}-\frac{2c_{W}}{s_{W}}\frac{\Pi^{\rm 1PI}_{Z\gamma}(0)}{m_{Z}^{2}}
-\frac{c_{W}^{2}}{s_{W}^{2}}\left(\frac{\Pi^{\rm 1PI}_{ZZ}(m_{Z}^{2})}{m_{Z}^{2}}-\frac{\Pi^{\rm 1PI}_{WW}(m_{W}^{2})}{m_{W}^{2}} \right) \\ 
 &+\frac{\Pi^{\rm 1PI}_{WW}(0) -\Pi^{\rm 1PI}_{WW}(m_{W}^{2})}{m_W^2}
+\frac{\alpha_{\rm em}}{4\pi s_W^2} \left(6+\frac{7-4s_W^2}{2s_W^2}\log c^2_W \right).
\end{align}}
{Using an expansion for $\Pi_{WW}^{\rm 1PI}(q^{2})$, $\Pi_{ZZ}^{\rm 1PI}(q^{2})$ at $q^{2}=0$, namely, 
\begin{align}
\label{eq:ex}
\frac{\Pi_{VV}^{\rm 1PI}(q^{2})}{m_{V}^{2}}=\frac{\Pi_{VV}^{\rm 1PI}(0)}{m_{V}^{2}}+\Pi_{VV}^{\rm 1PI}(0)^{\prime}\frac{q^{2}}{m_{V}^{2}}+...\ \ \ \ \ (V=W,Z),
\end{align}
above expression for $\Delta r$ can be {rewritten} as
\begin{align}
\label{eq:dran}
\Delta r &=
-\frac{c_{W}^{2}}{s_{W}^{2}}\Delta \rho
+\Delta \alpha_{\rm em}
+(\Delta r)_{\rm remainder}.
\end{align}
{where $\Delta \rho=\Pi^{{\rm 1PI}}_{ZZ}(0)/m_{Z}^{2}-\Pi^{{\rm 1PI}}_{WW}(0)/m_{W}^{2}$,
and $(\Delta r)_{\rm remainder}$ denotes remaining parts of $\Delta r$. 
We note that the $m_{f}^{2}$ and $m_{\phi}^{2}$ dependence only appear in $\Delta \rho$. 
In $\Delta \alpha_{\rm em}$ such quadratic dependence disappear due to the Ward identity (instead, large logarithmic dependence of light fermions appear). 
 In addition, $(\Delta r)_{\rm remainder}$ only includes higher order terms in the expansion Eq.~\eqref{eq:ex}, which do not provide quadratic mass terms as can be easily seen by dimensional analysis. }} }}

The NNLO QCD corrections to $h\to q\bar{q}$ are expressed in the ${\rm\overline{MS}}$ scheme in a limit neglecting contributions with quark masses  as~\cite{Djouadi:2005gi,Spira:2016ztx,Chetyrkin:1995pd}
\begin{align}
\notag
\Delta_{\text{QCD}}^q &= 5.67\frac{\alpha_s(m_h)}{\pi}+(35.94-1.36 N_f)\left(\frac{\alpha_s(m_h)}{\pi}\right)^2 \\ 
&+\frac{\kappa_t}{\kappa_q}\left(\frac{\alpha_s(m_h)}{\pi}\right)^2\left\{1.57-\frac{2}{3}\log\frac{m_h^2}{m_t^2}+\frac{1}{9}\log\left(\frac{\bar{m}^2_Q(m_h)}{m_h^2}\right)^2\right\},
\end{align}
where $N_f$ is the active flavor number and the renormalization scale is taken to be at the mass of the Higgs boson, $\mu=m_h$, in this expression. 
While the first and the second terms are common in the extended Higgs models and the SM, the third term, which comes from the top loop contributions at the NNLO, contains the ratio of scaling factors of Yukawa couplings 
$\kappa_t/\kappa_q$. 
When we apply the QCD corrections, we regard the quark mass as the running ${\rm \overline{MS}}$ mass in the Yukawa couplings appeared in Eq.~\eqref{eq:hff-lo}, which is also evaluated at $\mu=m_h$.
The running quark masses are calculated from the ${\rm \overline{MS}}$ mass at $\mu=m_q$ by using the relation~\cite{Gray:1990yh}, i.e., $\bar{m}_q(m_h)=\bar{m}_q(m_q)c[\alpha_s(m_h)/\pi]/c[\alpha_s(m_q)/\pi]$, 
where the function $c$ can be found up to the three loop level in Refs.~\cite{Gorishnii:1990zu,Gorishnii:1991zr,Chetyrkin:1997dh,Vermaseren:1997fq}.
In {{\tt H-COUP\_ver~2}}, { for purpose that users can perform numerical checks, }  two options for the computations of $\Gamma_0(h\to q\bar{q})$ can be selected: the one is computations with current masses for quarks, and the other is computations used ${\rm \overline{MS}}$ mass for the Yukawa couplings.

The decay rates for {$h\to VV^\ast \to Vf\bar{f}$} can be expressed as the same manner with $h\to f\bar{f}$, i.e.,
\begin{align}
\Gamma(h\to Vf\bar{f})=\Gamma_0(h\to Vf\bar{f})\left[1+\Delta_{\rm EW}^V+\Delta_{\rm QCD}^V\right]. 
\end{align}
Here the decay rate at the LO, $\Gamma_0(h\to Vf\bar{f})$,  is presented in terms of a fraction $\epsilon_V=m_V/m_h$ by~\cite{Keung:1984hn} 
\begin{align}
\Gamma(h\to Vf\bar{f})=\frac{\sqrt{2}G_fm_hC_V\left(\Gamma_{hVV}^{1,{\rm tree}}\right)^2}{768\pi^3}F(\epsilon_V),
\end{align}
where the factor $C_V$ is $C_V=4(v_f^2+a_f^2)$ for the Z boson and $C_V=1$ for the W boson, and the function $F(\epsilon_V)$ is written as
\begin{align}
\notag
F(\epsilon_V)&=\frac{3(1-8\epsilon_V^2+20\epsilon_V^4)}{\sqrt{4\epsilon_V^2-1}}{\rm arccos}\left(\frac{3\epsilon_V^2-1}{2\epsilon_V^3}\right) \\ 
&-(1-\epsilon_V^2)\left(\frac{47}{2}\epsilon_V^2-\frac{13}{2}+\frac{1}{\epsilon_V^2}\right)-3(1-6\epsilon_V^2+\epsilon_V^4)\log\epsilon_V.
\end{align}  
For $h\to Zf\bar{f}$, further separation of the EW correction $\Delta_{\rm EW}^V$ into the QED part and the weak part can be performed as $\Delta_{\rm EW}^Z =\Delta_{\rm QED}^Z+\Delta_{\rm Weak}^Z$, similar to $h\to f\bar{f}$. 
 The NLO QED correction $\Delta_{\rm QED}^Z$ is given by the same expression to the SM~\cite{Kniehl:1993ay} in the $m_f^{} \to 0$ limit; i.e,
 \begin{align}
 \Delta_{\rm QED}^Z=Q_f^2\frac{3\alpha_{\rm em}}{4\pi}, \label{eq:QED_hZff}
 \end{align}
since the QED corrections only appear in the vertex of the off-shell Z boson with a pair of fermions, which does not have new physics effects in the massless limit of decaying fermions. 
In contrast to $h\to Zf\bar{f}$, for $h\to Wf\bar{f}^\prime$, such separation cannot be done because the Feynman diagrams with a virtual photon are accompanied by virtual W bosons. 
The weak corrections to $h\to Zf\bar{f}$ and the EW corrections to $h\to Wf\bar{f^\prime}$, namely $\Delta_{\rm Weak}^Z$ and $\Delta_{\rm EW}^W$ are expressed in terms of renormalized vertices of $hVV$ and $hf\bar{f}$ as well as other contributions; e.g. oblique corrections to the off-shell weak boson and box diagrams for $h\to Vf\bar{f}$. 
The explicit formulae can be found in Ref.~\cite{Kanemura:2019kjg}. 
On the other hand, the NLO QCD correction to $h\to Vq\bar{q}$ in the $\rm{\overline{MS}}$ scheme is commonly presented by~\cite{Kniehl:1993ay}
\begin{align}
\Delta_{\rm QCD}^V=C_F\frac{3\alpha_s(m_h)}{4\pi},
\end{align}
with $C_F = 3/4$. 
In {{\tt H-COUP\_ver~2}} the three-body-decays of the Higgs boson $\Gamma(h\to Vf\bar{f})$ are implemented. However, four body decays $\Gamma(h\to 4f)$ are not included. They are calculated with NLO EW and NLO QCD corrections in HSM~\cite{Altenkamp:2018bcs} and THDMs~\cite{Altenkamp:2017kxk}.

The loop induced decays of the Higgs boson are also evaluated in {{\tt H-COUP\_ver~2}}, i.e., $\Gamma(h\to gg)$ and $\Gamma(h\to \mathcal{V} \gamma)\ (\mathcal{V}=Z,\gamma)$, including higher order QCD corrections. 
Analytical formulae for these processes can be found in Refs.~\cite{Kanemura:2015fra,Kanemura:2019kjg}, Refs.~\cite{Kanemura:2015mxa,Kanemura:2018esc,Kanemura:2019kjg}, and Refs.~\cite{Kanemura:2016sos,Kanemura:2018esc,Kanemura:2019kjg} for the HSM, THDMs and the IDM, respectively. 

For the $\Gamma(h\to gg)$, the QCD corrections up to NNLO in the ${\rm \overline{MS}}$ scheme are implemented in {{\tt H-COUP\_ver~2}}. 
The analytic expression for $m_h^2/m_t^2\to 0$ is taken~\cite{Djouadi:2005gi,Chetyrkin:1997iv},
\begin{align}
\Delta_{\rm QCD}^{g}=\frac{215}{12}\frac{\alpha_s(m_h)}{\pi}+\left(\frac{\alpha_s(m_h)}{\pi}\right)^2\left(156.8-5.7\log\frac{m_t^2}{m_h^2}\right),
\end{align}
where the active flavor number $N_f$ and the renormalization scale $\mu$ have been taken to be $N_f=5$ and $\mu=m_h$, respectively. 
Typically, the corrections of the NLO contribution  (the first term) and the NNLO contribution (the second terms) are about 70\% and 20\% to the LO contributions, respectively. 
For the $\Gamma(h\to \gamma\gamma)$, the QCD corrections up to NNLO are implemented in the limit $ m_t\to\infty$ in the program. 
In this process, the QCD corrections are only implemented to the top loop diagrams; because that to the another quark loop contributions are numerically negligible. 
Thus, top loop contributions denoted as $\big(\Gamma^{\rm loop}_{h\gamma\gamma}\big)_{t}$  are modified at the amplitude level as~\cite{Djouadi:2005gi,Steinhauser:1996wy}
\begin{align}
\big(\Gamma^{\rm loop}_{h\gamma\gamma}\big)_{t}  \to \big(\Gamma^{\rm loop}_{h\gamma\gamma}\big)_t\left[1-\frac{\alpha_s(\mu)}{\pi}-\Bigg(\frac{\alpha_s(\mu)}{\pi}\Bigg)^2\left(\frac{31}{24}+\frac{7}{4}\log\frac{\mu^2}{m_t^2}\right)\right],  \label{eq:hgamgam_QCD}
\end{align}
where we take $\mu=m_h/2$, following Ref.~\cite{Djouadi:2005gi}\footnote{In Ref.~\cite{Djouadi:2005gi}, validity for taking the renormalization scale at $\mu=m_h/2$ is also discussed.} . 
Apart from $h\to\gamma\gamma$, for $h\to Z\gamma$, only NLO corrections, which are given by the second terms in Eq.~\eqref{eq:hgamgam_QCD}, are applied in {{\tt H-COUP\_ver~2}}. 
Typical size of the NLO QCD corrections to the LO contributions is $\mathcal{O}(0.1)\%$, so that the NNLO corrections can be negligible.

\section{Structure of {{\tt H-COUP\_ver~2}}}\label{sec:structure}

\begin{figure}[t]
\begin{center}
\includegraphics[width=150mm]{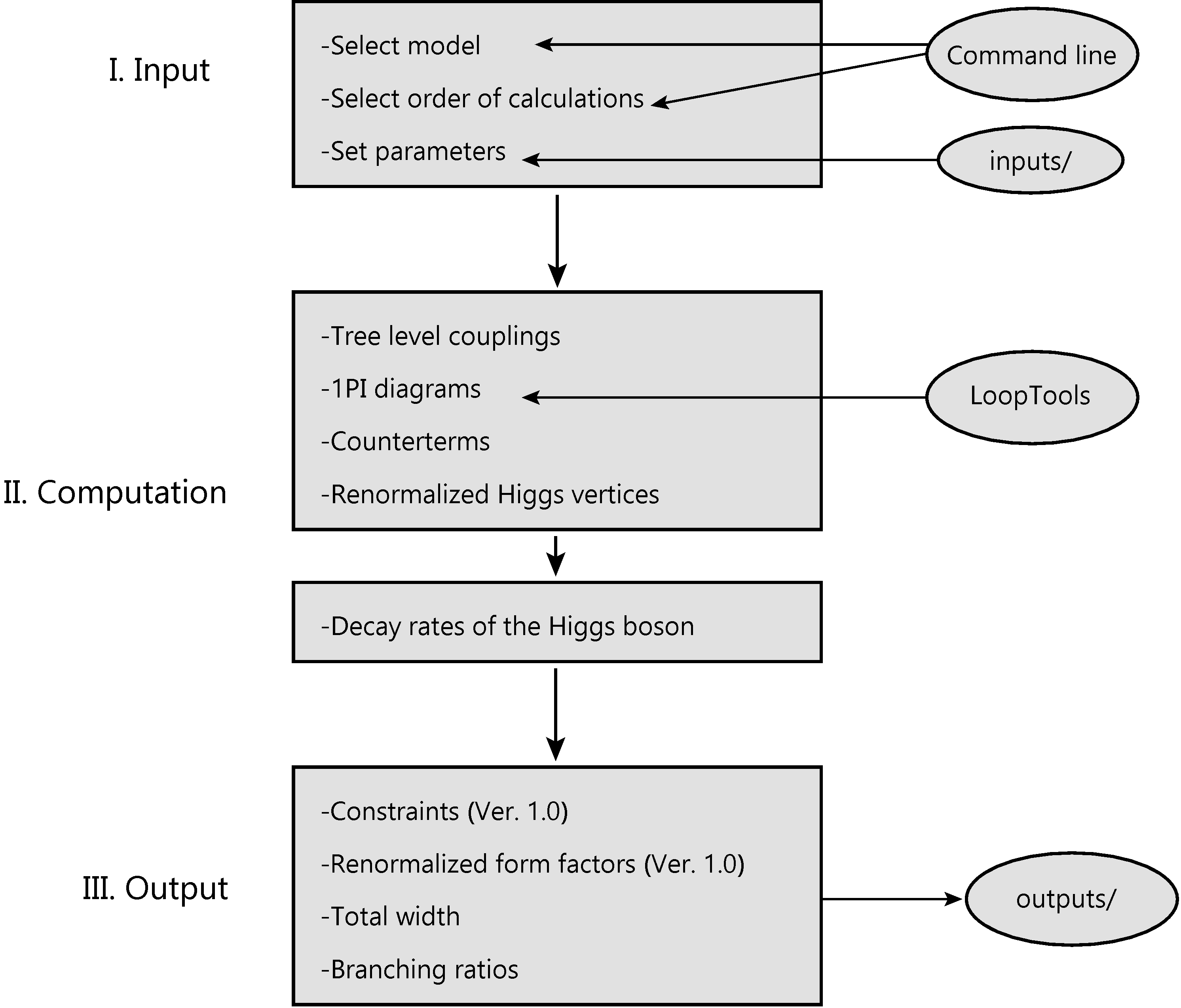}
\caption{Structure of {{\tt H-COUP\_ver~2}} }
\label{h2}
\end{center}
\end{figure}

The structure of {{\tt H-COUP\_ver~2}}  is schematically shown in Fig.~\ref{h2}. 
Differently from {\tt H-COUP\_1.0}, the model and the order of calculations are specified from the command line interface (see Sec.~\ref{sec:how}). 
{{\tt H-COUP\_ver~2}} then reads the model independent (global) and model dependent input parameters, where 
the former is the SM inputs and the squared momenta of the renormalized form factors, which are commonly used in all the model files. 
The SM parameters and their default values are summarized in Table~\ref{global1}. 
In this table, $\Delta\alpha_{\rm em}$ denotes the shift of the fine structure constant given at the zero energy  $\alpha_{\rm em}$ to that given at  the Z boson mass $\alpha_{\rm em}(m_Z^{})$, i.e., 
\begin{align}
\alpha_{\rm em}(m_Z^{}) = \frac{\alpha_{\rm em}}{1 - \Delta\alpha_{\rm em}}. 
\end{align}
The strong coupling constant $\alpha_s(m_Z^{})$ is given at the Z boson mass. 
For the calculation of the Higgs boson decay rates, we have to use the 
strong coupling constant at different energy scale $\mu$ such as the Higgs boson mass as discussed in Sec.~\ref{sec:decay}, which is calculated by using the RGE running {at the 3-loop~\cite{Tarasov:1980au,Larin:1993tp}}. 
For the bottom and charm quark masses, we show both the on-shell and $\overline{\rm MS}$ masses, where 
the former masses can be derived from the latter by perturbative calculations. 
In {{\tt H-COUP\_ver~2}}, we simply quote the value of these on-shell masses from~\cite{Tanabashi:2018oca}.

For the squared momenta, their input values are only used to output values of the renormalized form factors of the Higgs boson, so that 
users who are interested in the width and the branching ratios of the Higgs boson do not need to take care of these parameters. 
For details of the treatment of the squared momenta, see Ref.~\cite{Kanemura:2017gbi}. 
The model dependent parameters and their default values are summarized in Tables~\ref{input_hsm},~\ref{input_thdm} and \ref{input_idm} for the HDM, the THDMs and the IDM, respectively. 
{In these Tables, $\Lambda$ denotes a cut off scale that is relevant for the theoretical constraints mentioned in Sec.II, i.e., triviality bound and vacuum stability bound.}
Here, we note that in {\tt H-COUP\_1.0} the type of Yukawa interactions (Type-I, -II, -X and -Y) can be specified from the input file of the THDM, but in {{\tt H-COUP\_ver~2}} it
can now be specified from the command line interface. 
Therefore, the ``Type'' parameter in the THDMs in {\tt H-COUP\_1.0} disappears in {{\tt H-COUP\_ver~2}}. 

\begin{table}[t]
\begin{tabular}{|llll|}\hline
Parameter                           & Definition in {{\tt H-COUP\_ver~2}}  & Description                       & Default value   \\\hline \hline 
  $m_Z$                             & {\sf mz}                         &  $Z$ mass                         & 91.1876 GeV \\\hline 
  $\alpha_{\text{em}}$               & {\sf alpha\_em}                  & Fine structure constant           & 137.035999139$^{-1}$       \\\hline 
  $G_F$                             & {\sf G\_F}                       & Fermi constant                    & 1.1663787$\times 10^{-5}$ GeV$^{-2}$  \\\hline 
  $\Delta \alpha_{\text{em}}$        & {\sf del\_alpha}                 &  Shift of $\alpha_{\text{em}}$      & 0.06627                  \\\hline 
  $\alpha_s(m_Z^{})$                        &{\sf alpha\_s}                    & Strong coupling            & 0.1181              \\\hline 
  $m_h$                             & {\sf mh}                         & Higgs boson mass                  & 125.1 GeV            \\\hline            
  $m_t$                             & {\sf mt}                         & On-shell $t$ mass                 & 173.1 GeV   \\\hline
  $m_b$                             & {\sf mb}                         & On-shell $b$ mass                 & 4.78 GeV       \\\hline 
  $\bar{m}_b(m_b)$                  & {\sf mb\_ms}                     & $\overline{\rm MS}$ $b$ mass      & 4.18 GeV       \\\hline 
  $m_c$                             & {\sf mc}                         & On-shell $c$ mass                 & 1.67 GeV    \\\hline 
  $\bar{m}_c(m_c)$                  & {\sf mc\_ms}                     & $\overline{\rm MS}$ $c$ mass      & 1.27 GeV    \\\hline 
  $m_\tau$                          & {\sf mtau}                       & $\tau$ mass                       & 1.77686 GeV  \\\hline
  $m_\mu$                           & {\sf mmu}                       & $\mu$ mass                         & 0.1056583745 GeV  \\\hline
 \end{tabular}
\caption{Input global SM parameters. All these parameters are defined by double precision, and their input values are taken from particle data group~\cite{Tanabashi:2018oca}.}
\label{global1}
\end{table} 

\begin{table}[t]
\begin{tabular}{|c|| c c c c c c|}\hline
                  & \multicolumn{6}{c|}{HSM}     \\\hline\hline
Parameters        & $m_H^{}$ & $\alpha$  & $\mu_S^{}$ & $\lambda_S$ & $\lambda_{\Phi S}^{}$ & $\Lambda$ \\\hline 
{\tt H-COUP} def. & {\sf mbh} & {\sf alpha} & {\sf mu\_s}  & {\sf lam\_s} & {\sf lam\_phis} & {\sf cutoff} \\\hline
Default value     & 500 GeV     & 0.1    & 0    & 0.1     & 0     & 3 TeV     \\\hline
\end{tabular}
\caption{Input parameters in the HSM.  All these parameters are defined by double precision.  }
\label{input_hsm}\vspace{5mm}
\begin{tabular}{|c||  c c c c c c c c|}\hline
                  & \multicolumn{8}{c|}{THDM}     \\\hline\hline
Parameters        &  $m_{H^\pm}^{}$ & $m_A^{}$ & $m_H^{}$ & $M^2$ & $s_{\beta-\alpha}$ & $\text{Sign}(c_{\beta-\alpha})$  & $\tan\beta$ & $\Lambda$\\\hline 
{\tt H-COUP} def. &  {\sf mch}          & {\sf ma}      & {\sf mbh}      & {\sf bmsq}  & {\sf sin\_ba}                  & {\sf sign} ($+1$ or $-1$)   & {\sf tanb} & {\sf cutoff}\\\hline 
Default value     &  500 GeV     & 500 GeV      & 500 GeV       & (450 GeV)$^2$ & 1 & 1 & 1.5 & 3 TeV        \\\hline
\end{tabular}
\caption{Input parameters in the THDMs. 
All these parameters are defined by double precision except for {\sf sign} which is defined by integer, and can be either 1 or $-1$. }
\label{input_thdm}\vspace{5mm}
\begin{tabular}{|c||c c c c cc|}\hline
                  & \multicolumn{6}{c|}{IDM}     \\\hline\hline
Parameters        &  $m_{H^\pm}^{}$ & $m_A^{}$ & $m_H^{}$ & $\mu_2^2$ & $\lambda_2$ & $\Lambda$ \\\hline 
{\tt H-COUP} def. &  {\sf mch}    & {\sf ma}      & {\sf mbh}      &  {\sf mu2sq}    & {\sf lam2}        & {\sf cutoff}  \\\hline 
Default value     &  500 GeV      & 500 GeV       & 500 GeV       &  (500 GeV)$^2$   & 0.1                 & 3 TeV \\\hline
\end{tabular}
\caption{Input parameters in the IDM. All these parameters are defined by double precision.  }
\label{input_idm}
\end{table}

In the computation block, tree-level Higgs boson couplings, 1PI diagrams and counterterms are calculated under the fixed model and input parameters. 
These calculations are then used to compute the decay rates of the Higgs boson. 

In the output block, {{\tt H-COUP\_ver~2}} tells us if a given configuration determined by the input parameters is allowed or excluded. 
If a given parameter choice is excluded, a message ``Excluded by XXX'' appears, where ``XXX'' can be perturbative unitarity, vacuum stability, triviality, true vacuum conditions and/or ST parameters. 
In the both cases, the output file is generated in the output directory. 
{{\tt H-COUP\_ver~2}} provides the decay branching ratios and the total width as well as outputs given in {\tt H-COUP\_1.0} (the renormalized form factors). 



{In {\tt H-COUP\_ver~2}, implemented decay processes are fixed as those of the 125 GeV Higgs boson in the SM. 
Since {\tt H-COUP\_ver~2} does not include the decay into extra Higgs bosons, {a case for such non-standard processes cannot be applied}. }

\section{Installation and how to run\label{sec:how}}

In order to run the {\tt H-COUP} program, users need to install a Fortran compiler (GFortran is recommended) and {\tt LoopTools}~\cite{Hahn:1998yk} in advance. 
One can download the {\tt LoopTools} package from \cite{Hahn:1998yk}, and see the manual for its installation. 

One can download the {\tt H-COUP} program on the following webpage\\[-1cm]
\begin{verbatim}
    http://www-het.phys.sci.osaka-u.ac.jp/~hcoup
\end{verbatim}
In the following, we describe how to run {{\tt H-COUP\_ver~2}} in order. 

\begin{enumerate}
\item Unzip the {HCOUP-2.X.zip} file:
\begin{center}
\fbox{
 \$  unzip {HCOUP-2.X.zip}}
\end{center}
Then, the {HCOUP-2.X} directory ({HCOUP-2.X/}) is created.
In this directory, one can find 3 files (Makefile, README, main.F90) and 4 directories
as follows:
\begin{center}
\fbox{
\begin{tabular}{l}
 \$ ls \\
 Makefile README main.F90 inputs/ models/ modules/ outputs/ 
\end{tabular}
}
\end{center}
Each directory contains the following files:
\begin{itemize}
\item inputs/ (input files for the model dependent/global parameters)\\ 
 in\_hsm.txt      (input file for the HSM) \\
 in\_thdm.txt     (input file for the THDMs) \\
 in\_idm.txt      (input file for the IDM)  \\
 in\_sm.txt       (global input file for the SM parameters)  \\
 in\_momentum.txt (global input file for momenta)   
\item outputs/ (output files generated by {\tt H-COUP})\footnote{Initially this directory is empty.}\\
 out\_hsm.txt, outBR\_hsm.txt (output files for the HSM) \\
 out\_thdm.txt, outBR\_thdm.txt (output files for the THDMs) \\
 out\_idm.txt, outBR\_idm.txt (output files for the IDM)  \\
 out\_sm.txt, outBR\_sm.txt (output files for the SM) 
\item models/ (main Fortran90 files of {\tt H-COUP}) \\
 HCOUP\_HSM.F90 (main file for the HSM)  \\
 HCOUP\_THDM.F90 (main file for the THDMs)\\
 HCOUP\_IDM.F90 (main file for the IDM) 
\item modules/ (module files of {\tt H-COUP}) 
\end{itemize}
We note that users do not need to touch the files in models/ and modules/, but  
only need to modify the input files in inputs/.

\item Open Makefile by an editor and replace ``PATH\_TO\_LOOPTOOLS'' appearing in the line 
  ``LPATH''   
   by the correct path to the library file of {\tt LoopTools} (libooptools.a). 

\item To compile the code, execute
 \begin{center}
 \fbox{\$ make}
 \end{center}
  in the {HCOUP-2.X} directory. Then, an executable file ``hcoup'' is generated. 

\item To run the {\tt H-COUP} program, execute
 \begin{center}
 \fbox{\$ ./hcoup}
 \end{center}
   Then, you are asked,\\[-10mm]
\begin{verbatim}
Which model? (1=HSM, 2=THDM-I, 3=THDM-II, 4=THDM-X, 5=THDM-Y, 6=IDM)
\end{verbatim}
\vspace*{-4mm}
   in the command line. You can specify the model by typing the number.
   You are further asked,\\[-10mm]
\begin{verbatim}  
Which order for EW? (0=LO, 1=NLO)
\end{verbatim}    
\vspace*{-4mm}
   and \\[-10mm]
\begin{verbatim}    
Which order for QCD? (-1=LO(quark mass:OS), 0=LO(quark mass:MSbar), 1=NLO, 
2=NNLO)
\end{verbatim}    
\vspace*{-4mm}   
   in order. 
   You can specify the order of calculations by typing the numbers, see also Sec.~\ref{sec:decay-rate} for details of quark masses. 

   Then, output files are generated in the output directory. 
   If a given set of the input parameters is excluded by some of the constraints, 
   a message appears in the command line. 
   An example of the generated output file in outputs/ is shown in Fig.~\ref{file_out}. 

\item One can change the model-dependent input parameters by modifying the in\_hsm.txt, in\_thdm.txt and in\_idm.txt files in the input directory.
One can also change the SM parameters and the squared momenta of the renormalized Higgs vertices by modifying the in\_sm.txt and in\_momentum.txt files in the input directory. 
These two files are commonly used to all the model files for each extended Higgs model.
In Fig.~\ref{file_in}, we show an example of the input file for the HSM (in\_hsm.txt). 
  
\end{enumerate} 

As a physics example computed by {{\tt H-COUP\_ver~2}}, we also present branching ratios of the SM-like Higgs boson 
in four types of THDMs in Fig.~\ref{br}, where the NLO-EW and NNLO-QCD corrections are taken into account.

\begin{figure}
\fontsize{11pt}{9pt}
\begin{verbatim}
BLOCK MODEL #
     1      1   # HSM
BLOCK BSMINPUTS #
     1      1.00000000E-01   # alpha
     2      0.00000000E+00   # lambda_{phi S}
     3      1.00000000E-01   # lambda_S
     4      0.00000000E+00   # mu_S (GeV)
BLOCK SMINPUTS #
     1      7.29735257E-03   # alpha_em
     2      1.16637870E-05   # Fermi constant
     3      1.18100000E-01   # alpha_s
     4      1.27000000E+00   # mc(mc) MSbar
     5      4.18000000E+00   # mb(mb) MSbar
     6      1.67000000E+00   # mc On-shell
     7      4.78000000E+00   # mb On-shell
BLOCK MASS #
     4      5.66262421E-01   # mc(mh) MSbar
     5      2.79078561E+00   # mb(mh) MSbar
     6      1.73100000E+02   # mt
    13      1.05658374E-01   # mmu
    15      1.77686000E+00   # mtau
    23      9.11876000E+01   # mz
    24      8.09388629E+01   # mw (calculated,tree)
    24      8.04132574E+01   # mw (calculated,1-loop)
    25      1.25100000E+02   # mh
    35      5.00000000E+02   # mH
BLOCK CONSTRAINTS #
     0      3.00000000E+03   # The cutoff scale (GeV)
     1      0   # Vacuum stability at tree level [0=OK, 1=No]
     2      0   # Tree-level unitarity [0=OK, 1=No]
     3      0   # S and T parameters [0=OK, 1=No]
     4      0   # True vacuum [0=OK, 1=No]
     5      0   # Vacuum stability (RGE improved with the cutoff scale) [0=OK, 1=No]
     6      0   # Triviality (with the cutoff scale) [0=OK, 1=No]
#
# Decay width of the SM-like Higgs boson by H-COUP #
#          PDG          Width
DECAY       25     0.38906463E-02   # EW:NLO QCD:NNLO                
#          BR         NDA       ID1      ID2
     2.56864747E-02     2         4       -4    # BR(h -> c c~)
     6.02006318E-01     2         5       -5    # BR(h -> b b~)
     2.26520896E-04     2        13      -13    # BR(h -> mu- mu+)
     6.52833529E-02     2        15      -15    # BR(h -> tau- tau+)
     8.16713494E-02     2        21       21    # BR(h -> g g)
     2.38864676E-03     2        22       22    # BR(h -> gam gam)
     1.64408658E-03     2        22       23    # BR(h -> gam Z)
     2.26679585E-02     2        23       23    # BR(h -> Z Z*)
     1.98425293E-01     2        24      -24    # BR(h -> W+ W-*)
\end{verbatim}
\caption{Example of the output file (outBR\_hsm.txt)} 
\label{file_out}   
\end{figure}  

\begin{figure}
\fontsize{11pt}{9pt}
\begin{verbatim}
!================================!
!                                !
! Input parameters for the HSM   !
!                                !
!================================!
  500.d0           ! m_H in GeV
  0.1d0            ! alpha
  0.d0             ! lambda_{phi S}
  0.1d0            ! lambda_S
  0.d0             ! mu_S in GeV
  3.d3             ! cutoff in GeV
\end{verbatim}
\caption{Example of the input file (in\_hsm.txt)} 
\label{file_in}   
\end{figure}  

\begin{figure}
 \includegraphics[width=0.4\textwidth]{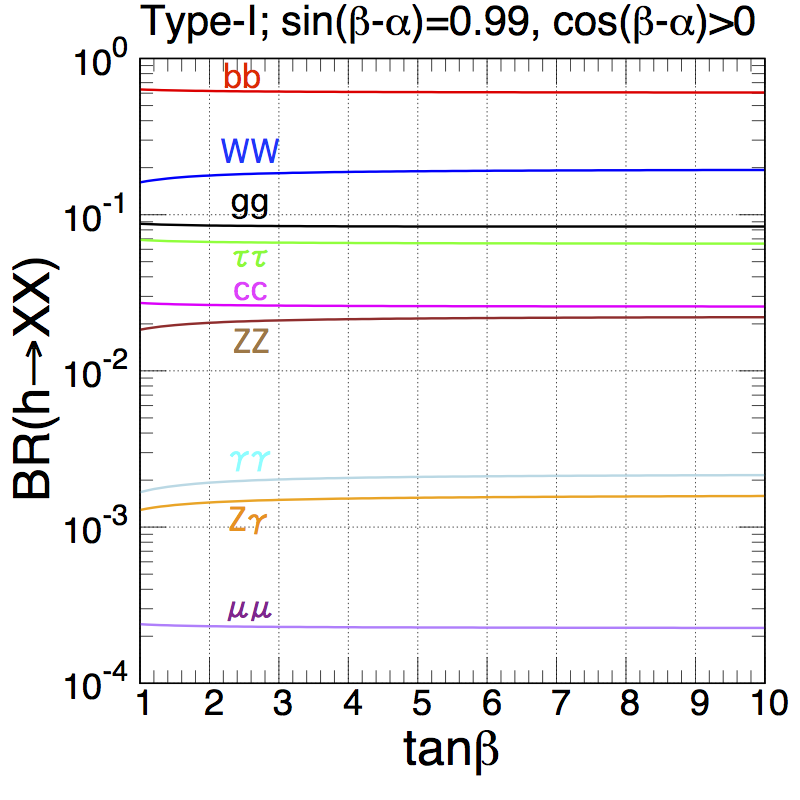}\quad
 \includegraphics[width=0.4\textwidth]{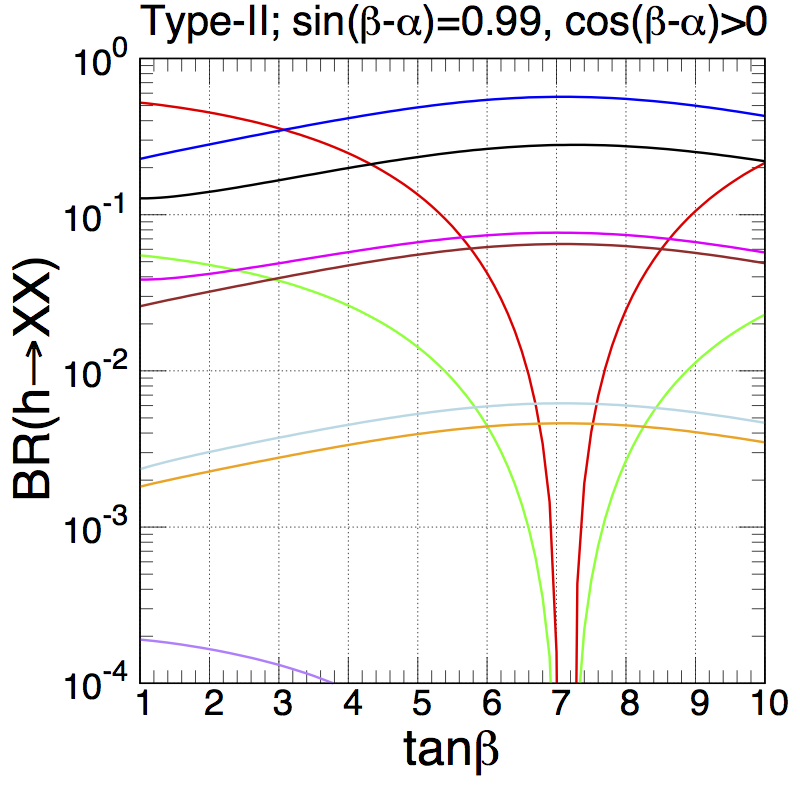}\\[2mm]
 \includegraphics[width=0.4\textwidth]{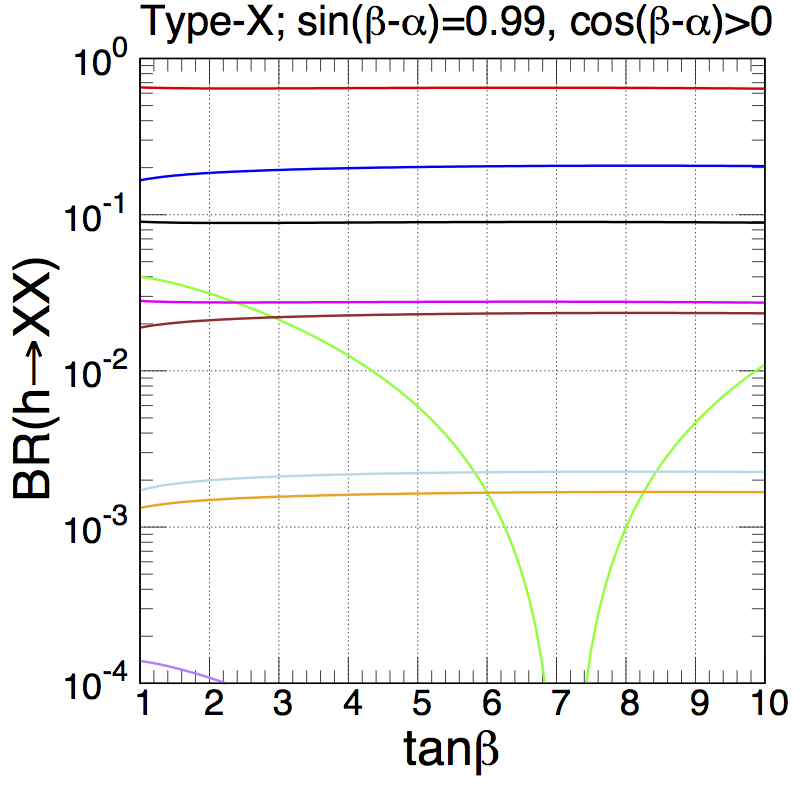}\quad
 \includegraphics[width=0.4\textwidth]{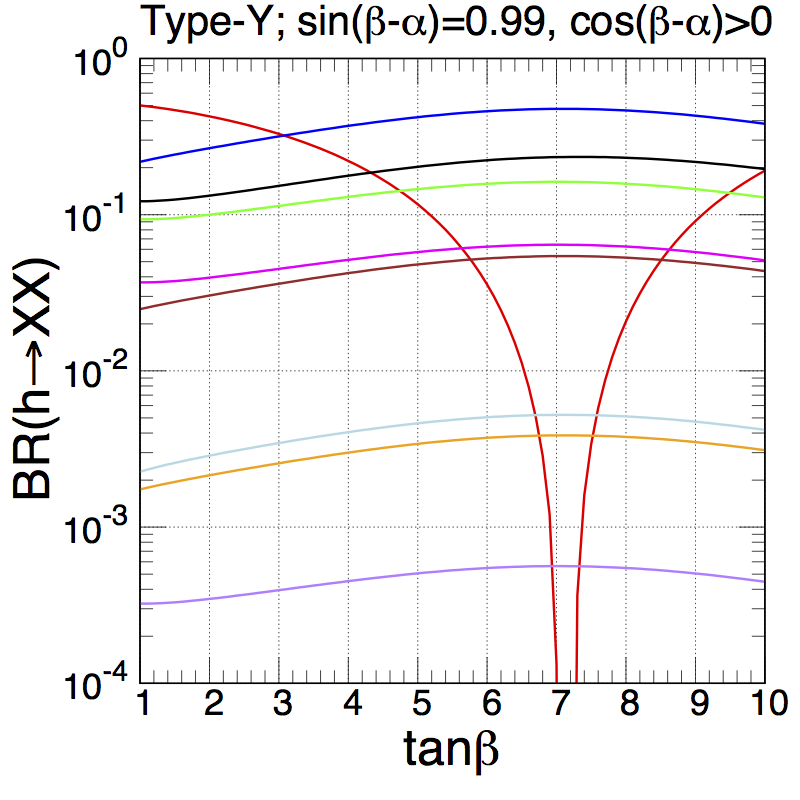}   
\caption{Branching ratios of the 125-GeV Higgs boson as a function of $\tan\beta$ in the Type-(I, II, X, Y) THDMs
for $\sin(\beta-\alpha)=0.99$ with $\cos(\beta-\alpha)>0$, where we take $M=m_H=300$~GeV and $m_A=m_{H^\pm}=600$~GeV.} 
\label{br}   
\end{figure}

\section{summary}\label{sec:summary}

In this paper, the concept and the manual of {{\tt H-COUP\_ver~2}} have been presented, 
which is a set of fortran programs for numerical evaluation of decay rates of the Higgs boson with a mass of 125 GeV  
and the decay width with higher order corrections (NLO for EW and scalar loop corrections, and NNLO for QCD corrections) for  
various models of extended Higgs sectors. 
In {{\tt H-COUP\_ver~2}}, in addition to the SM, 
the Higgs singlet model, four types of two Higgs doublet models with a softly-broken $Z_2$ symmetry and the inert doublet model are implemented. 
{{\tt H-COUP\_ver~2}} contains all the functions of {\tt H-COUP\_1.0} where a full set of the Higgs boson vertices are evaluated at one-loop level 
in a gauge invariant manner in these models. 
We have briefly introduced these models with their theoretical and experimental constraints, and we have summarized formulae for the renormalized vertices and the decay rates. 
After the explanation of the structure of the program, we have demonstrated how to install and run {{\tt H-COUP\_ver~2}} with some numerical examples.

\begin{acknowledgments}
This work is supported in part by the Grant-in-Aid on Innovative Areas, the Ministry of Education, Culture, 
Sports, Science and Technology, No.~16H06492 and No.~18H04587 [S.K.],  
JSPS KAKENHI Grant No.~18K03648 [K.M.], 
JSPS KAKENHI Grant No.~18J12866 [K.S.],
and Early-Career Scientists, No.~19K14714 [K.Y.]. 
\end{acknowledgments}

\bibliography{bibhcoup}

\end{document}